\documentclass[aip,amsmath,amssymb,preprint]{revtex4-1}
\usepackage{graphicx}
\usepackage{bm}
\newcommand{\be}{\begin{equation}}
\newcommand{\ee}{\end{equation}}
\newcommand{\bea}{\begin{eqnarray}}
\newcommand{\eea}{\end{eqnarray}}
\draft

\begin{document}

%\preprint{AIP/123-QED}

\title{Exact partition function zeros and the collapse transition of a two-dimensional lattice polymer}

\author{Jae Hwan Lee}
\affiliation{School of Systems Biomedical Science\\and\\Department of Bioinformatics and Life Science,\\Soongsil University, Seoul 156-743, Korea}

\author{Seung-Yeon Kim}
\affiliation{School of Liberal Arts and Sciences, Chungju National University, Chungju 380-702, Korea}

\author{Julian Lee}
\email{jul@ssu.ac.kr}
\affiliation{School of Systems Biomedical Science\\and\\Department of Bioinformatics and Life Science,\\Soongsil University, Seoul 156-743, Korea}
%\affiliation{School of Systems Biomedical Science and Department of Bioinformatics and Life Science, Soongsil University, Seoul 156-743, Korea}
\affiliation{Department of Pharmacuetical Chemistry and Graduate Group in Biophysics, University of California, San Francisco 94158, USA}
\date{\today}

\begin{abstract}

We study the collapse transition of the lattice homopolymer on a square lattice
by calculating the exact partition function zeros. The exact partition function is obtained by enumerating
the number of possible conformations for each energy value, and the exact distributions of the partition function zeros
are found in the complex temperature plane by solving a polynomial equation.
We observe that the locus of zeros closes in on the positive real axis as the chain length increases, providing the evidence for the onset of the collapse transition. By analyzing the scaling behavior of the first zero with the polymer length, we estimate the transition temperature $T_\theta$ and the crossover exponent $\phi$.
%, whose values are consistent with results of earlier works.
\end{abstract}



\maketitle

%%%%%
\section{Introduction}
%%%%%
The hydrophobic interaction and the excluded volume effect are two
main interactions that determine the conformation of a polymer in a
dilute solution, in space dimension $d<4$~\cite{CD91}.
In the good solvent regime, the repulsive excluded volume effect is the
dominating factor and the mean end-to-end distance
$R_N$ of a polymer chain with $N$ monomers asymptotically grows as
$\langle R_N^2 \rangle \sim N^{6/(d+2)}$, the behavior of a
self-avoiding random walk. On the other hand, the poor solvent
regime is defined by the property that the attractive hydrophobic
interaction between monomers dominates, where the scaling behavior is
$\langle R_N^2 \rangle \sim N^{2/d}$. The situation is usually
described by the statement that the polymer adopts a swollen
conformation in a good solvent and the collapsed one in a poor
solvent. The solvent where the repulsive and attractive interactions
cancel each other is called the theta solvent, with the
corresponding temperature being called the theta temperature, or
Flory temperature, $T_\theta$~\cite{F49}. $T_\theta$ is the
temperature where the condition of the solvent changes from good to
poor or \textit{vice versa} and the collapse transition occurs. The collapse transition has been studied and the critical exponents have been calculated using various theoretical and computational methods~\cite{F67,dG75,dG78,dG79,S75,B82,KF84,BBE85,DS85,P86,S86,DS87,SS88,PCJS89,ML89,
CM93,GH95,BBG98,NKMR01,ZOZ08,CDC09,GV09}, including lattice models~\cite{B82,BBE85,DS85,P86,S86,DS87,SS88,PCJS89,ML89,
CM93,GH95,BBG98,NKMR01,ZOZ08,CDC09,GV09}. In particular, the self-avoiding walk on a square lattice has been extensively
studied as a model for the polymer in two dimensions,
and its collapse transition has been studied using exact
enumeration~\cite{DS85,S86} and Monte Carlo
samplings~\cite{B82,BBE85,SS88,PCJS89,ML89,CM93,GH95,BBG98,NKMR01,ZOZ08,CDC09,GV09}.

Alternatively, phase transitions can be studied by calculating partition function zeros. The study of partition function zeros was initiated by the seminal paper of Yang and Lee~\cite{YL52},
where the zeros in the complex fugacity plane were studied to give a new
insight on the phase transition. Subsequently the zeros in the
complex temperature plane were studied by Fisher~\cite{F65}. With the
recent advance of computational power, the study of partition
function zeros became one the most popular methods for studying the
phase transition and critical phenomena~\cite{BDL05}, and was used to study helix-coil transition of poly-alanine~\cite{AH00} and folding transition of
 a simple model protein~\cite{WW03}.
 However, it was rarely used for the study of lattice polymers, although some preliminary qualitative results
on collapse transition were reported for both homopolymers~\cite{L04} and heteropolymers~\cite{WW03,CL05}.

The power of the partition function zeros method lies in its
sensitivity to the onset of a phase transition. When the energy takes discrete values, the partition function $Z$ is expressed as
\be
Z = \sum_E n(E) e^{-\beta E}
\ee
with $n(E)$ being the number of states with energy $E$.
When $Z$  is a function of
$y \equiv e^{\beta \epsilon}$ with some interaction parameter $\epsilon$, such as when $E$ values are integer multiples of $\epsilon$,
the partition function can be expressed in the form
\be
Z(y)  = A(y) \prod_{i} (y-y_i)
\ee
where $A(y)$ is a function which is analytic in the whole complex plane, and $y_i$s are the complex roots of the equation $Z(y)=0$, called the partition function zeros. Since $Z$ is real, $y_i$s form conjugate pairs except for the real-valued ones.
By taking log and derivatives,
one obtains the specific heat
\be
C_N (T) = \frac{k_B}{N}(\ln y)^2 \left[ \sum_{i} \left\{\frac{y}{y-y_i} - \left( \frac{y}{y-y_i} \right)^2\right\}+ \left( y \frac{d}{d y} \right)^2 \ln A(y) \right]
\label{sh}
\ee
where $N$ is the size parameter of the system such as the particle number.
For a system with the phase transition at $y=y_c$, the locus of the zeros close in toward the positive real axis to intersect it at $N=\infty$,
and the singularity of $C_N(T)$ appears in this limit.
It is clear from Eq.~(\ref{sh}) that the leading behavior of such a singularity is due to the pair of partition function zeros closest to the real axis, called the first zeros.
Therefore, by calculating the partition function zeros and examining the behavior of the first zeros as $N \rightarrow \infty$,   the critical behavior can be much more accurately analyzed than examining the behavior of $C_N (T)$ for real values of the temperature, which is plagued by
   the noise due to the subleading terms containing zeros other than the first ones.

In this work,
we calculate the exact partition function zeros of the polymers on the square lattice up to the length $N=36$, and make extrapolation of the first zero positions
to estimate the collapse transition temperature $T_\theta$ and the crossover exponent $\phi$
(See the next section for the definition).
The fact that our calculation is exact, along with the sensitivity of the partition function zeros method,
allows us to estimate these quantities with reasonably high accuracy.
%We find that our results are consistent with most of the results reported earlier.

%%%%%%%%%%%%%%%%%%%%%%%%%%
\section{The Scaling Behavior and the Critical Exponent}
%%%%%%%%%%%%%%%%%%%%%%%%%%
The collapse transition is described by the scaling behavior of $R_N$ near the critical temperature~\cite{dG79},
\begin{equation}
\langle R_N^2 \rangle \sim N^{2\nu} f(\tau N^\phi), \label{scaling}
\end{equation}
where $\tau \equiv (T-T_\theta) / T_\theta$ is the reduced temperature, and $f(x)$ is a function with the property
\begin{eqnarray}
f(0) &=& 1 \nonumber \\
f(x) &=& \left\{
              \begin{array}{ll}
                   x^{\mu_+} & (x \rightarrow \infty)\\
                   x^{\mu_-} & (x \rightarrow -\infty),
              \end{array}
       \right.
\end{eqnarray}
with $\mu_\pm$ being exponents that reproduce the scaling behavior of $R_N^2$ in the good and poor solvent regime,
\bea
\mu_+ &=& \frac{6/(d+2)-2\nu}{\phi} \nonumber \\
\mu_- &=& \frac{2/d - 2\nu}{\phi}.
\eea
In most of the studies on lattice models, the transition temperature and the critical exponents were usually obtained by examining the behavior of $\langle R_N^2 \rangle$ as a function of
$N$ and $T$ and fitting to Eq.~(\ref{scaling}), but they could also be
obtained from the scaling behavior of the specific
heat~\cite{ML89,CM93}:
\be
C_N(T) \sim N^{\alpha \phi} g(\tau N^\phi), \label{sh2}
\ee
with
\begin{eqnarray}
g(x) &=& \left\{
              \begin{array}{ll}
                   A^+x^{-\alpha} & (x \rightarrow \infty)\\
                   {\rm const} & (x=0)\\
                   A^-x^{-\alpha} & (x \rightarrow -\infty),
              \end{array}
       \right.
\end{eqnarray}
The crossover exponent $\phi$ measures how rapidly the system undergoes the transition as the temperature approaches the critical temperature $T_\theta$. As will be shown later, it is directly related to the exponent that measures how rapidly the first zeros approach the positive real axis as $N \rightarrow \infty$.

%%%%%%%%%%%%%%%%%%%%%%%%%%
\section{The Model}
%%%%%%%%%%%%%%%%%%%%%%%%%%

A conformation of a polymer chain with $N$
monomers is modeled as  a two-dimensional self-avoiding chain of
length $N$ on a square lattice. The position of the monomer $i$ is given by ${\bf r}_i = (k,l)$,
where integers $k$ and $l$ are the Cartesian coordinates relative to an arbitrary origin.
Chain connectivity requires $\vert {\bf r}_i-{\bf r}_{i+1} \vert = 1$, i.e., bond length is unity.
Due to the excluded volume, there can be no more than one monomer on each lattice site,
${\bf r}_i \neq {\bf r}_j$ for $i \neq j$. The attractive hydrophobic interaction is incorporated by assigning
the energy $-\epsilon <0$ for each non-bonded contact between monomers. The resulting Hamiltonian is:
\begin{equation}
    {\cal H} = -\epsilon \sum_{i<j} \Delta ({\bf r}_i, {\bf r}_j),
\end{equation}
where
\be
\Delta ({\bf r}_i, {\bf r}_j) = \left\{
              \begin{array}{ll}
                  1 & (|i-j| > 1 \quad {\rm and} \quad |{\bf r}_i-{\bf r}_j| =1)\\
                   0 & ({\rm otherwise}).
              \end{array}
       \right.
\ee
Since the energy of the system is $E=-\epsilon K$ where $K$ is the number of monomer-monomer contacts,
 the partition function is expressed as a polynomial:
 \be
 Z = \sum_{K=0}^{K_{\rm max}(N)} \Omega_N(K) y^K,
 \ee
where $y  \equiv \exp(\beta \epsilon) $, $\Omega_N(K)$ the number of
polymer conformations with contact number $K$, and $K_{\rm max}(N)$
is the maximum number of possible contacts, when polymer length is
$N$~\cite{CD89}:
\begin{equation}
    K_{\max}(N) = \left\{
    \begin{array}{ll}
      N-2m & {\rm for} ~~ m^2 < N \le m(m+1),\\
      N-2m-1 ~~~ & {\rm for} ~~ m(m+1) < N \le (m+1)^2 ,
    \end{array} \right.
\end{equation}
where $m$ is a positive integer.
Therefore the partition function zeros can be obtained by enumerating
 the number of conformations $\Omega_N(K)$ for each contact number $K$.
The speed of enumeration can be increased by calculating the reduced number of conformations $\omega_N(K)$, where conformations related by rigid rotations, reflections, and translations are regarded as equivalent
and counted only once. However, it is assumed
that there is an intrinsic direction in the chain, so the
conformations related by the exchange of labels $i \leftrightarrow
N-i+1$ for all $(i=1,\cdots, N)$ are considered distinct. We note
that since the rigid rotations and reflections in two dimensions
form an eight-fold symmetry, the total number of conformations
generated by rotations and reflections from a given
two-dimensional conformation is eight. An exception is the
straight chain,  a one-dimensional conformation invariant with respect to reflection perpendicular to the
chain. Consequently, the total number of conformations generated
by rotations and reflections is four in this case.
 Therefore, the number of conformations with rigid rotations and reflections considered distinct,
 denoted by $\Omega_N(K)$,
can be easily obtained by
\bea
\Omega_N (0)&=&8\omega_N (0)-4 \nonumber\\
 \Omega_N (K)&=&8\omega_N (K) \quad (K>0).
 \label{total}
\eea
Using a parallel algorithm that classifies each conformation according to the size of box it spans~\cite{LKL10_1},
 we could calculate $\omega_N(K)$ up to $N=36$. The same quantities were calculated up to $N=28$ in earlier
  works~\cite{CD89,L04}, which agree with the current results.

%%%%%
\section{partition function zeros}
\begin{figure}
\includegraphics[width=.9\textwidth]{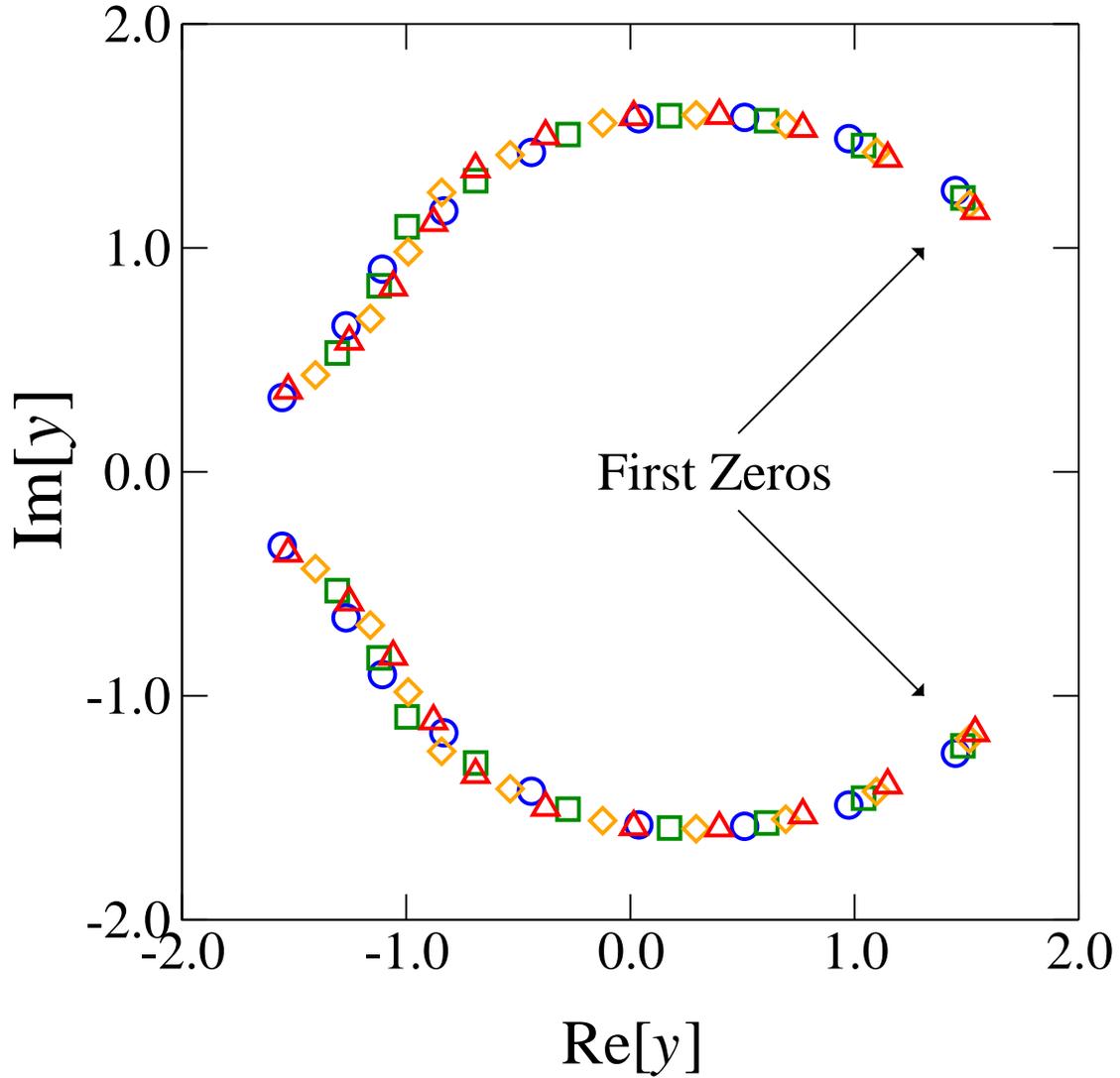}
\caption{Positions of the partition function zeros in the complex temperature ($y = e^{\beta \epsilon}$) plane
for $N=30$(circles), 32(squares), 34(diamonds), and 36(triangles).
The first zeros are the ones closet to the positive real axis.}
\label{fig1}
\end{figure}

%%%%%
The partition function zeros were obtained by solving the polynomial equation
\be
 Z = \sum_{K=0}^{K_{\rm max}(N)} \Omega_N(K) y^K = 0,
\ee
using \textsc{MATHEMATICA}.
We observe the partition function zeros fall on a simple locus, more or less independent of the polymer length $N$ (Fig.~\ref{fig1}).

\begin{figure}
\includegraphics[width=.9\textwidth]{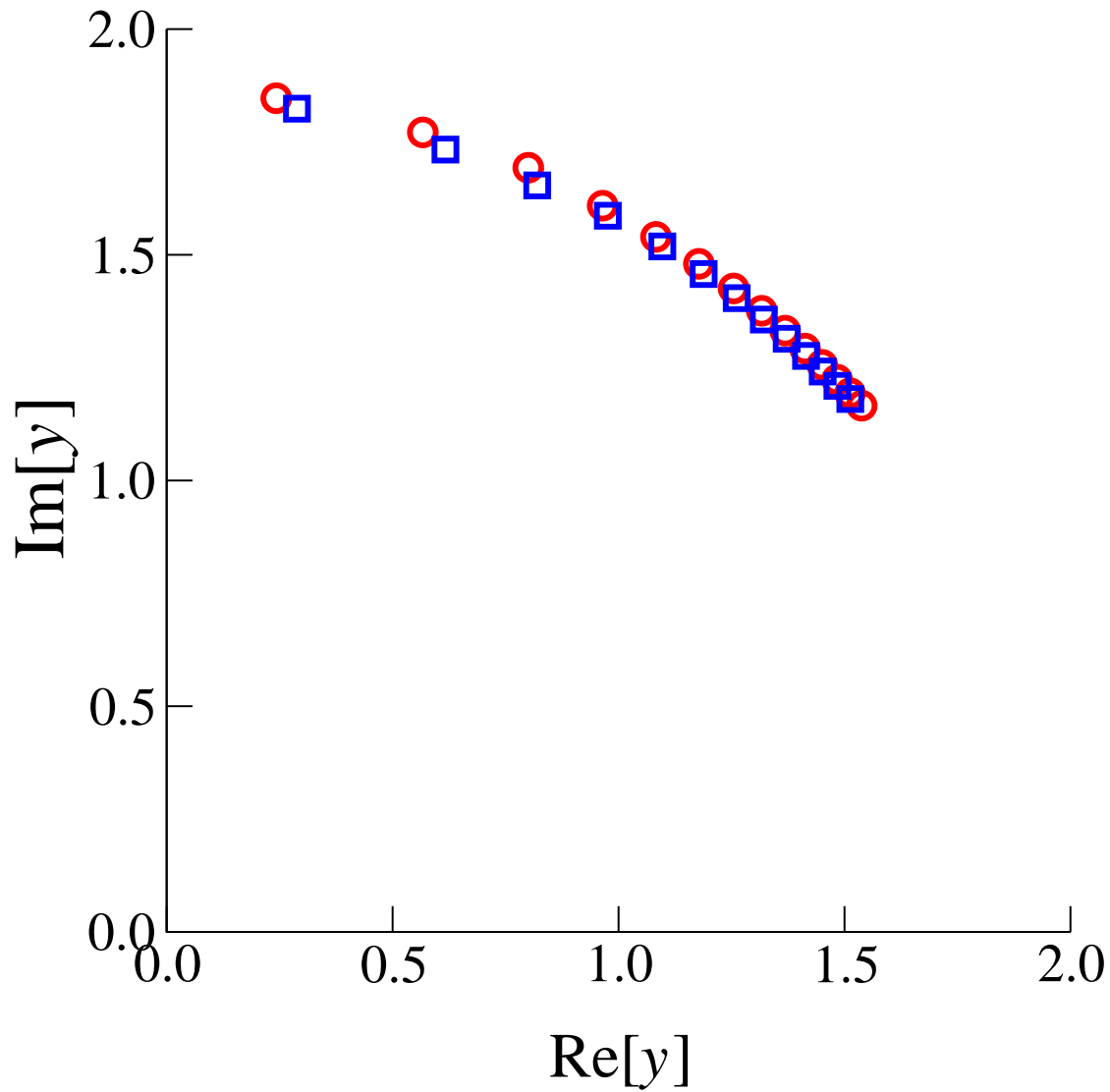}
\caption{Positions of the first zeros in the first quadrant of the complex temperature ($y = e^{\beta \epsilon}$) plane
for even lengths $N=10$, 12, 14, $\cdots$, 36 (circles) and
for odd lengths $N=11$, 13, 15, $\cdots$, 35 (squares) from left to right.
The first zeros approach the positive real axis as $N$ increases.}
\label{fig2}
\end{figure}

\begin{figure}
\includegraphics[width=.9\textwidth]{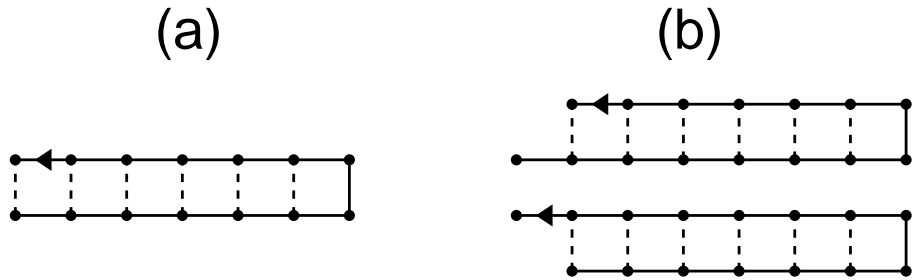}
\caption{The hairpin as an example of the class of conformations whose number depends crucially on the parity of $N$. There is only one conformation for even $N$ (a), whereas there are two possible  conformations for odd $N$ (b). Note that there is an instrinsic direction in a chain, indicated by an arrow. The dashed lines indicate the inter-monomer contacts.}
\label{hairpin}
\end{figure}

 Although there is a relatively large gap between the positive real axis and the first zeros by visual inspection (Fig.~\ref{fig1}), the first zeros approach the positive real axis (Fig.~\ref{fig2}), and the transition temperature and the crossover exponent can be calculated from their behavior in the $N \to \infty$ limit. However, an oscillatory behavior is observed,  due to the fact that there are classes of conformations whose numbers depend crucially on the parity of $N$. For example, there is only one hairpin conformation when $N$ is even, but there are two possible conformations for odd $N$(Fig.~\ref{hairpin}). Therefore $N$s for each parity are used separately when $N \rightarrow \infty$ limit is taken, so that the large error due to the oscillatory behavior is eliminated.
 %Similar parity effects are observed in other models such as Ising model in terms of lattice size.  The parity effect in terms of  the periodic length of the lattice was also observed for polymer~\cite{DS85}, which is irrelevant for the current work since the size of the polymer is fixed and the lattice size is infinite.

The crossover exponent $\phi$ can be obtained from examining how fast the first zeros approach the positive real axis as $N$ increases~\cite{IPZ83}.
From the scaling relation Eq.~(\ref{sh2}), we see that the partition function scales as
\be
\ln Z_N(\tau) \sim N^{\alpha \phi - 2 \phi}  g(\tau N^\phi)
\ee
and the equation for the first zero in the first quadrant
\be
Z(\tau_1) = 0
\ee
is invariant with changing $N$ only if
\be
\tau_1 \sim N^{-\phi}, \label{first}
\ee
which is related to the corresponding complex temperature $t_1$ as
\be
\tau_1 \equiv \frac{t_1-T_\theta}{T_\theta}
\ee
In terms of $t_1$, Eq.~(\ref{first}) is reexpressed as:
\be
t_1  \sim T_\theta + {\rm const} \cdot N^{-\phi} \label{first_t}
\ee
which is asymptotically equivalent to
\be
y_1  \sim y_c + {\rm const} \cdot N^{-\phi} \label{first_y}
\ee
in the large $N$ limit, where $y_1= e^{\epsilon/t_1}$ and $y_c= e^{\epsilon/T_\theta}$.
From the imaginary part of Eq.~(\ref{first_y})
\begin{equation}
    \mathrm{Im}[y_1(N)] \sim N^{-\phi},
\label{im}
\end{equation}
the finite-size approximation of the crossover exponent is obtained:
\begin{equation}
    \phi(N) = - \frac{\ln\{{\rm Im}[y_1(N+2)]/{\rm Im}[y_1(N)]\}}{\ln\{(N+2)/N\}}. \label{bst}
\end{equation}
%where $N$ for each parity is used separately in order to eliminate the large error due to the oscillatory behavior.
The expression Eq.~(\ref{bst}) reduces to the exact value of $\phi$
in the $N \to \infty$ limit, which is estimated by using the
Bulirsch-Stoer (BST) extrapolation~\cite{BS64,PTVF92,M02}. For given
$m$ data points corresponding to distinct values of $N$, the BST
extrapolation is performed by constructing a rational function of
$(1/N)^\omega$ that passes through all of these points, under the
assumption that the leading finite-size correction is of order
$O((1/N)^\omega)$. Then, the extrapolated value is obtained by
evaluating the function at $1/N=0$. The estimated error is defined
as~\cite{HS88,AFH00,M00}
\be 2 | \phi_{-1} - \phi_{-m} | \ee
where $\phi_{-i}$ is the
 value of $\phi$ at $1/N=0$ extrapolated from the data with the $i$-th point eliminated. The estimated error is the
measure for the robustness of the extrapolated value with respect to
perturbations in the data points, but it has no statistically
rigorous confidence level associated with it.  The estimated error
can be further reduced by removing unreliable data obtained from
$N<20$, and the final result is \be \phi = 0.422(12), \ee obtained
from the data for even $N$ with $20 \le  N \le 36$. In the absence
of additional information, we assumed the leading finite size
correction to $\phi$ is of order $O(N^{-1})$ when performing the BST
procedure, but extrapolated value of $\phi$ does not seem to depend
much on this assumption(data not shown).

Once the value of $\phi$ is determined, the transition temperature $T_\theta$ can be obtained by estimating the point on the positive real axis where the first zeros approach in the limit of $N \rightarrow \infty$, applying the BST extrapolation procedure to the real part of
Eq. (\ref{first_y}),
\begin{equation}
    \mathrm{Re}[y_1(N)]-y_c \sim N^{-\phi}.
\label{ac}
\end{equation}
 The resulting value of $y_c$ is
$2.16(18)$, equivalent to $T_\theta/\epsilon = 1.30(17)$, where again, the data for even $N$
with $20 \le  N \le 36$ were used.

\begin{figure}
\includegraphics[width=.9\textwidth]{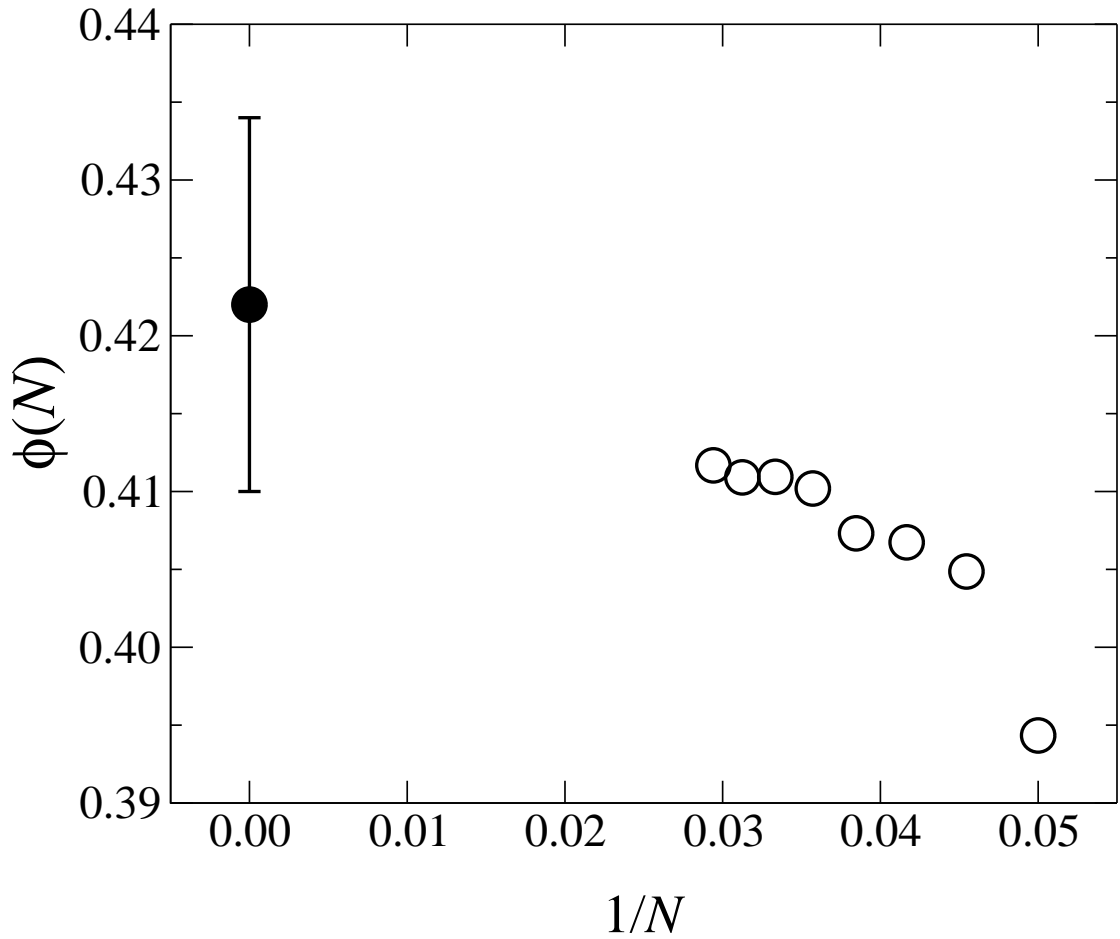}
\caption{The finite size approximations of the crossover exponent, $\phi(N)$, are shown
as a function of $1/N$ for even $N$ with $N \ge 20$ (open circles),
and the value of $\phi$ at infinite size obtained by the BST extrapolation is indicated by a solid circle with an error bar.}
\label{phi}
\end{figure}

\begin{figure}
\includegraphics[width=.9\textwidth]{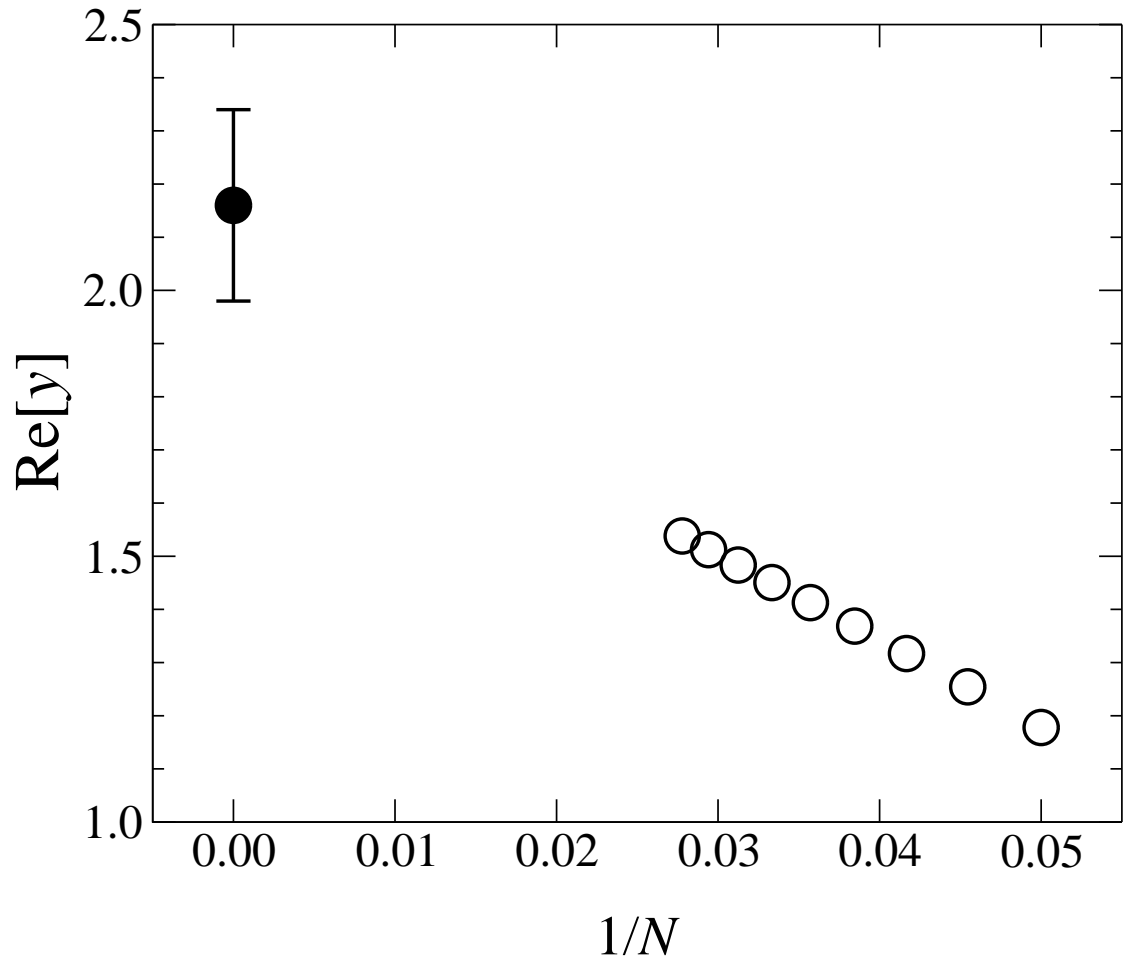}
\caption{Values of the real part of the first zeros are shown as a function of $1/N$ for even $N$ with $N \ge 20$ (open circles), and the value for $N \rightarrow \infty$ obtained by the BST extrapolation is indicated by a solid circle with an error bar.}
\label{Tth}
\end{figure}

The finite value approximations of $\phi$ and $y_c$ are displayed in Figs.~\ref{phi} and \ref{Tth} as functions of $1/N$, along with their extrapolated values at $1/N=0$.  The extrapolated value of $y_c$ in Fig.~\ref{Tth} is larger than obtained by drawing a straight line through the data points, because we assumed the leading behavior of $y-y_c$ being proportional to $(1/N)^{0.422}$. There is no change of the extrapolated value of $T_\theta$ under the current precision when we use the conjectured exact value of the crossover exponent $\phi=3/7$~\cite{DS87} instead of $\phi=0.422$. The values obtained in the current study are compared with those from the
earlier works in Table~\ref{table}.
Since $T_\theta/\epsilon$ is not a universal quantity, it is displayed only for the square lattice polymer with nearest neighbor interaction.
The maximum sizes of the polymer studied, $N_{\max}$, are displayed wherever applicable. The results of the current study are given in the
first line of the Table~\ref{table}. Although there are variations in the results reported earlier, we find that many of them are consistent with ours. Those that agree with our results within the estimated errors are indicated by boldface letters. In particular, it should be noted that the value of $\phi$ obtained in the current work agrees quite well with the exact value 3/7 obtained by analytic calculation on the polymers on the hexagonal lattice~\cite{DS87}, which is believed to be in the same universality class as those on the square lattice\cite{GH95,NKMR01,CDC09,GV09}.

%%%%%%%%%%%%%%
\section{Discussion}
%%%%%%%%%%%%%%

We studied the zeros of the exact partition function of lattice polymers
on square lattices up to chain length 36 by exhaustively enumerating the number of all possible conformations.
We observed that the first zeros tend to approach the positive real axis as the chain length increases,
and estimated the critical temperature $T_\theta$ and the crossover exponent $\phi$
by the BST extrapolation.

In contrast to Monte Carlo approaches where the calculation can be done for  polymer lengths up to several hundreds or thousands,
 the chain length studied in the current study is much shorter, but the exactness of our data allows us to use powerful extrapolation methods, leading to a reasonably accurate estimation of the transition temperature and the crossover exponent. Furthermore, by  studying the complex zeros of the partition function zeros, instead of examining the scaling behavior of real-valued quantities such as radius of gyration or specific heat, much more accurate analysis of the phase transition could be performed.

It is of immediate interest to perform the exact enumeration of polymer conformations up to sizes where the approach of the first zeros toward the positive real axis is more visible. An exact enumeration has been performed using a transfer matrix for length up to 72  at infinite temperature~\cite{J04}, and it would be interesting to see whether it can be generalized to count the number of conformations for each energy without introducing too much extra computational costs, in order to calculate the partition function zeros. One could also combine Monte Carlo methods with the partition function zeros to increase the polymer size, at the cost of introducing sampling error. There are indications that the locations of the first zeros are robust with respect to the sampling errors, a point that needs further investigation~\cite{LKL10}.

As a final remark, the partition function zeros method may be applied to study the transition behavior of heteropolymers\cite{WW03,CL05}, related to the very important and interesting topic of protein folding. In contrast to homopolymers, the definition of large $N$ limit is not so clear for a heteropolymer, so the finite size scaling argument such as the one used in the current study cannot be applied directly. Various methods to extract information relevant to the collapse and the folding transition, from the complex partition function zeros, will have to be explored.
%%%%%%%%%%%%%
\begin{acknowledgments}
This work was supported by Mid-career Researcher Program through NRF grant funded by the MEST (No.2010-0000220).
\end{acknowledgments}

%%%%%%%%%%%%

\newpage

\begin{table}
\caption{\label{table}The critical temperature $T_\theta$ and the crossover exponent $\phi$ obtained in the current work, displayed in the first line, are  compared with those in the literature.
$T_\theta$ is displayed only for the model of the current work. The results that agree with ours within the estimated errors are indicated by boldface letters. }
\begin{ruledtabular}
\begin{tabular}{cccll}

Method & lattice & $N_{\max}$ & ~ $T_\theta/\epsilon$ & ~~ $\phi$ \\

\hline

Exact partition function zeros & square & 36 & {\bf 1.30(17)} & {\bf 0.422(12)} \\
Field theory\cite{S75} & N/A & N/A & ~~~- &  ${7\over11}~(\approx 0.64)$\\
Renormalization group\cite{KF84} & N/A & N/A & ~~~- & ${19\over22}~(\approx 0.86)$\\
Monte Carlo~\cite{B82} & square & 160 & {\bf 1.31(6)} & ~~~- \\
Monte Carlo~\cite{BBE85} & square & 200 & {\bf 1.55(15)} & 0.6(1) \\
Transfer matrix~\cite{DS85,S86} & square & N/A & {\bf 1.42(4)} & {\bf 0.48(7)} \\
Series expansion~\cite{P86} & triangular & 16 & ~~~- & 0.64(5) \\
Coulomb gas method ~\cite{DS87}& hexagonal & N/A & ~~~- & $\bf {3\over7}~(\approx 0.43$) \\
Monte Carlo and renormalization group ~\cite{SS88} & square & 40  & {\bf 1.54(7)} & 0.52(7) \\
Monte Carlo~\cite{PCJS89} & hexagonal & 300 & ~~~- & {\bf 0.5(1)} \\
Scanning simulation~\cite{CM93} & square  & 240 & 1.52(1) & 0.530(4) \\
Recursive enrichment method~\cite{GH95} & square & 2048 & 1.504(5) & {\bf 0.435(6)} \\
The pruned-enriched Rosenbluth method~\cite{BBG98} & square & 256 & 1.4993(23) & ~~~- \\
%Analytic lattice~\cite{TL98} &  & 400(?) & ~~~- & 0.82\\
Interacting growth walk~\cite{NKMR01} & square & 2000 & ~~~-  & {\bf 0.419(3)} \\
Monte Carlo~\cite{ZOZ08} & square & 1600 & 1.50 & 0.545(4) \\
Monte Carlo~\cite{CDC09} & square & 300 & 1.505(18) & ~~~- \\
Monte Carlo~\cite{GV09} &  square\footnote{A model with explicit solvent molecules. Different from the model studied in this work.} & 20 &  ~~~- & {\bf 0.436(7)} \\

\end{tabular}
\end{ruledtabular}
\end{table}

\begin{thebibliography}{}
%%%%%%%%%%%%

\bibitem{CD91} H. S. Chan and K. A. Dill,
    Annu. Rev. Biophys. Biophys. Chem. \textbf{20}, 447 (1991).
\bibitem{F49} P. J. Flory, J. Chem. Phys. \textbf{17}, 303 (1949).
\bibitem{F67} P. J. Flory, \emph{Principles of Polymer Chemistry}
    (Cornell University Press, Ithaca, 1967).
\bibitem{dG75} P.-G. de Gennes, J. Physique Lett. \textbf{36}, 55 (1975).
%\bibitem{dG76} P.-G. de Gennes, J. Phys. France \textbf{37}, 1445 (1976).
\bibitem{dG78} P.-G. de Gennes, J. Physique Lett. \textbf{39}, 299 (1978).
\bibitem{dG79} P.-G. de Gennes, \emph{Scaling Concepts in Polymer Physics}
    (Cornell University Press, Ithaca, 1979).
\bibitem{S75} M. J. Stephen, Phys. Lett. \textbf{53}, 363 (1975).
\bibitem{KF84} A. L. Kholodenko and K. F. Freed, J. Phys. A \textbf{17}, L191 (1984);
    J. Chem. Phys. \textbf{80}, 900 (1984).
\bibitem{B82} A. Baumg\"artner, J. Physique \textbf{43}, 1407 (1982).
\bibitem{BBE85} T. M. Birshtein, S. V. Buldyrev, and A. M. Elyashevitch,
    Polymer \textbf{26}, 1814 (1985).
\bibitem{DS85} B. Derrida and H. Saleur, J. Phys. A \textbf{18}, L1075 (1985).
\bibitem{S86} H. Saleur, J. Stat. Phys. \textbf{45}, 419 (1986).
\bibitem{P86} V. Privman, J. Phys. A \textbf{19}, 3287 (1986).
\bibitem{DS87} B. Duplantier and H. Saleur, Phys. Rev. Lett. \textbf{59}, 539 (1987).
\bibitem{SS88} F. Seno and A. L. Stella, J. Phys. France \textbf{49}, 739 (1988).
\bibitem{PCJS89} P. H. Poole, A. Coniglio, N. Jan, and H. E. Stanley,
    Phys. Rev. B \textbf{39}, 495 (1989).
%\bibitem{ML89} H. Meirovitch and H. A. Lim, Phys. Rev. Lett. \textbf{62}, 2640 (1989);
%    J. Chem. Phys. \textbf{91}, 2544 (1989).
\bibitem{CM93} I. Chang and H. Meirovitch, Phys. Rev. E \textbf{48}, 3656 (1993).
\bibitem{GH95} P. Grassberger and R. Hegger, J. Phys. I France \textbf{5}, 597 (1995).
%\bibitem{TL98} M. P. Taylor and J. E. G. Lipson, J. Chem. Phys. \textbf{109}, 7583 (1998).
\bibitem{BBG98} G. T. Barkema, U. Bastolla, and P. Grassberger, J. Stat. Phys. \textbf{90}, 1311 (1998).
\bibitem{NKMR01} S. L. Narasimhan, P. S. R. Krishna, K. P. N. Murthy, and M. Ramanadham,
    Phys. Rev. E \textbf{65}, 010801(R) (2001).
\bibitem{ZOZ08} J. Zhou, Z.-C. Ou-Yang, and H. Zhou, J. Chem. Phys. \textbf{128}, 124905 (2008).
\bibitem{CDC09} A. G. Cunha-Netto, R. Dickman, and A. A. Caparica,
    Comput. Phys. Comm. \textbf{180}, 583 (2009).
\bibitem{GV09} M. Gaudreault and J. Vi\~nals, Phys. Rev. E \textbf{80}, 021916 (2009).
\bibitem{ML89} H. Meirovitch and H. A. Lim, Phys. Rev. Lett. \textbf{62}, 2640 (1989);
    J. Chem. Phys. \textbf{91}, 2544 (1989).
\bibitem{YL52} C. N. Yang and T. D. Lee, Phys. Rev. \textbf{87}, 404 (1952);
    T. D. Lee and C. N. Yang, Phys. Rev. \textbf{87}, 410 (1952).
\bibitem{F65} M. E. Fisher, in {\it Lectures in Theoretical Physics},
    edited by W. E. Brittin (University of Colorado Press, Boulder, 1965), Vol. 7c, p. 1.
%%%
\bibitem{BDL05} I. Bena, M. Droz, and A. Lipowski, Int. J. Mod. Phys. B \textbf{19}, 4269 (2005) and references therein.
\bibitem{AH00} N. A. Alves and U. H. E. Hansmann, Phys. Rev. Lett. \textbf{84}, 1836 (2000); Physica A \textbf{292}, 509 (2001).
\bibitem{WW03} J. Wang and W. Wang, J. Chem. Phys. \textbf{118}, 2952 (2003).


%%%

\bibitem{L04} J. Lee, J. Korean Phys. Soc. \textbf{44}, 617 (2004).
\bibitem{CL05} C.-N. Chen and C.-Y. Lin, Physica A \textbf{350}, 45 (2005).
\bibitem{CD89} H. S. Chan and K. A. Dill, Macromolecules \textbf{22}, 4559 (1989).
\bibitem{LKL10_1} Parallel Algorithm for Calculation of the Exact Partition Function of a Lattice Polymer, J. H. Lee, S.-Y. Kim, and J. Lee, Comput. Phys. Commun. (2011) (in press).

%%% scaling
%\bibitem{EKB82} E. Eisenriegler, K. Kremer, and K. Binder, J. Chem. Phys. \textbf{77}, 6296 (1982).

%%% partition function zeros

\bibitem{IPZ83} C. Itzykson, R. B. Pearson, and J. B. Zuber, Nucl. Phys. B \textbf{220}, 415 (1983).

%%% BST
\bibitem{BS64} R. Bulirsch and J. Stoer, Numer. Math. \textbf{6}, 413 (1964).
\bibitem{PTVF92} W. H. Press, S. A. Teukolsky, W. T. Vetterling, and B. P. Flannery,
    Numerical Recipes in C, 2nd edition (Cambridge University Press, Cambridge, 1992), p. 111.
\bibitem{M02} J. L. Monroe, Phys. Rev. E \textbf{65}, 066116 (2002).
\bibitem{HS88} M. Henkelt and G. Sch\"utz, J. Phys. A  \textbf{21} 2617
(1988).
\bibitem{AFH00} N. A. Alves, J. R. D. de Felicio, and U. H. E. Hansmann, J. Phys. A \textbf{33}, 7489 (2000).
\bibitem{M00} J. L. Monroe, Phys. Rev. E \textbf{64}, 016126 (2000).
%\bibitem{HP87} M. Henkel and A. Patkos, J. Phys. A \textbf{20}, 2199 (1987).
\bibitem{J04} I. Jensen, J. Phys. A \textbf{37}, 5503 (2004).
\bibitem{LKL10} J. H. Lee, S.-Y. Kim, and J. Lee, in preparation.


\end{thebibliography}
\end{document}